\documentclass[english,aps,prd,
groupedaddress,amssymb,
showpacs,nofootinbib]{revtex4}

\usepackage{times}
\usepackage{amsmath}
\usepackage{graphicx}

\newcommand{\beq}{\begin{eqnarray}}
\newcommand{\eeq}{\end{eqnarray}}

\makeatletter
\numberwithin{equation}{section}
\numberwithin{figure}{section}

\makeatother

\usepackage{babel}

\begin{document}
\setlength{\unitlength}{1mm}

\title{Longitudinal Rescaling of Quantum Chromodynamics}

\author{Axel \surname{Cort\'es Cubero}}

\email{acortes_cubero@gc.cuny.edu}

\author{Peter \surname{Orland}}

\email{orland@nbi.dk}

\affiliation{1. Baruch College, The 
City University of New York, 17 Lexington Avenue, 
New 
York, NY 10010, U.S.A. }

\affiliation{2. The Graduate School and University Center, The City University of New York, 365 Fifth Avenue,
New York, NY 10016, U.S.A.}

\begin{abstract}
We examine the effect of quantum longitudinal rescaling of coordinates, {\em i.e.} 
$x^{0,3}\to\lambda x^{0,3}$,  $x^{1,2}\to x^{1,2}$ on the action of quantum chromodynamics (with quarks) to one loop. We use an 
aspherical Wilsonian integration (previously applied to the pure Yang-Mills theory and to quantum electrodynamics). Quantum fluctuations produce anomalous powers of $\lambda$ in the coefficients of the rescaled action. Our results are valid for small rescalings only, because perturbation theory breaks down if $\lambda\ll1$. 
\end{abstract}


\maketitle

\section{Introduction}

An important kinematic regime of quantum chromodynamics (QCD) is at very high center-of-mass energies and small transverse 
momenta. This kinematic regime is of major importance both at RHIC and the LHC.

A field-theory action can be simplified by a longitudinal rescaling of coordinates of the form $x^{0,3}\to\lambda x^{0,3}$,  $x^{1,2}\to x^{1,2}$ and $p_{0,3}\to\lambda^{-1}p_{0,3},\,p_{1,2}\to p_{1,2}$. The center-of-mass energy squared transforms as $s\to\lambda^{-2}s$ under this rescaling. The high-energy limit is $\lambda\rightarrow0$ \cite{Verlinde-squared}, 
\cite{McLerranVenugopalan}. The transverse degrees of freedom become unimportant, yielding a simpler effective action. 

Verlinde and Verlinde \cite{Verlinde-squared} argued that the longitudinally rescaled action yields BFKL theory \cite{BFKL}. McLerran and Venugopalan used a similar idea to derive the Color-Glass Condensate picture  \cite{McLerranVenugopalan}.

In Reference \cite{nearintegrability}, the longitudinal rescaling was used to express the anisotropic Yang-Mills theory as a set of coupled $(1+1)$-dimensional integrable field theories. It was shown that the theory has a mass gap and confines quarks in this limit and appears to have some of the correct features of soft-scattering. 

It is simple to see how such a rescaling affects the action of QCD. Non-Abelian gauge fields transform as $A_{0,3}\to\lambda^{-1}A_{0,3}$, and $A_{1,2}\to A_{1,2}$. The Yang-Mills action then transforms as
\beq
S_{\rm gauge}=\frac{1}{4g_{0}^{2}} \int d^{4}x F^{\mu\nu}F_{\mu\nu} \to\frac{1}{2g_0^2}\int d^4x {\rm Tr}\left(\lambda^{-2}F_{03}^2+\sum_{j=1}^2F_{0j}^2-\sum_{j=1}^2F_{j3}^2-\lambda^2F_{12}^2\right),\label{classicalrescaling}
\eeq
where $F_{\mu\nu}=\partial_\mu A_\nu-\partial_\nu A_\mu+i[A_\mu,A_\nu]$. The transverse electric and magnetic field strengths wildly fluctuate in the high-energy limit \cite{nearintegrability}, 
\cite{PhysRevD80}. The quark field $\psi$, transforms as 
$\psi\to\lambda^{\frac{1}{2}}\psi$. The action for this field becomes
\beq
S_{\rm Dirac}\to i\,\int d^4x\,\bar{\psi}\left(\gamma^0D_0+\gamma^3D_3+\lambda\gamma^1D_1+\lambda\gamma^2D_2\right)\psi,\label{classicalrescalingquark}
\eeq
where $D_\mu=\partial_\mu-iA_\mu$ is the covariant derivative. The effect of the rescaling is to suppress transverse transport of color and enhance
longitudinal transport. This can be most easily seen in the 
Hamiltonian formalism \cite{nearintegrability}, \cite{PhysRevD80} .

The rescaling (\ref{classicalrescaling}) and (\ref{classicalrescalingquark}) is purely classical. The 
$\lambda \rightarrow0$ limit is different in the quantum theory. Anomalous powers of 
$\lambda$ appear in the action, due to quantum fluctuations. The energy scale is changed by a longitudinal rescaling, so we require a renormalization procedure to tell us how the couplings change.

To understand how longitudinal-rescaling works in quantum field theory, consider the functional integral with an ultraviolet
lattice cut-off $a$. Longitudinal rescaling changes the spacing in the $0$- and $3$-directions to 
$\lambda a$, but in the $1$- and $2$-directions the spacing remains $a$. This 
yields a theory with an anisotropic cut-off. Quantum longitudinal rescaling therefore requires two steps. First we perform a Kadanoff or ``block-spin" transformation, by which some degrees of freedom are integrated out, leaving an renormalized theory on a lattice with spacing $\lambda^{-1}a$ in the longitudinal directions and spacing $a$ in the transverse directions. Then we apply the rescaling, which restores the isotropy; as a result, there is only one lattice spacing, rather than two. We refer the reader to Reference \cite{PhysRevD80} for more discussion. In practice, 
real-space renormalization procedures are quite difficult. We therefore 
use Wilson's momentum-space renormalization procedure \cite{WK} with ellipsoidal sharp momentum cut-offs to integrate out the high-momentum degrees of freedom.

The procedure we discuss has been applied to pure Yang-Mills theory \cite{PhysRevD80}\footnote{There were some errors in the coefficients of the longitudinally-rescaled pure-Yang-Mills action in Reference \cite{PhysRevD80}.  These are corrected here.} and to quantum electrodynamics \cite{qed}.

The next section is a review of Wilsonian renormalization. We review standard results with spherical momentum cut-offs in Section III. We generalize to the aspherical cut-off case in section IV and find the QCD action after integrating out high-momentum degrees of freedom in section V. In Section VI, we present the longitudinally-rescaled action. We discuss the implications of the result in Section VII.

\section{Wilsonian Renormalization}

This section is a quick review of Wilsonian renormalization for the QCD action, and contains no new results. The Euclidean QCD action is
\beq
S=\int d^4x\left( \frac{1}{4g_0^2}\,F_{\mu\nu}^a\,F^{a\,\mu\nu}+\bar{\psi} \,/\!\!\!\!D \psi\right),\nonumber
\eeq
where we have used $A_{\mu}=A_{\mu}^{a}t^{a}$ where $t^a$ are the $SU(N)$ generators, 
with $a=1,2,...N^2-1$, normalized as ${\rm Tr}\,t^{a}t^{b}=\delta^{ab}$. The field strength components
are 
and $F_{\mu\nu}=F_{\mu\nu}^{a}t^{a}$, where
$F_{\mu \nu}^{a}=\partial_\mu A_{\nu}^{a}-\partial_{\nu} A_{\mu}^{a}
+if^{abc}A_{\mu}^{b}, A_{\nu}^{c}$, and where $f^{abc}$ are the structure coefficients. The slash on a vector quantity $J_\mu$ is $/\!\!\!J=\gamma^\mu J_\mu$, where 
$\gamma^\mu$ are the Euclidean Dirac matrices.

We choose real positive cut-offs $\Lambda$ and $\tilde \Lambda$, with units of
$cm^{-1}$ and two dimensionless real positive anisotropy parameters $b$ and $\tilde b$. These quantities must satisfy $\Lambda>{\tilde \Lambda}$, $b\ge 1$, ${\tilde b}\ge 1$ 
and $\Lambda^{2}/b \ge {\tilde \Lambda}^{2}/{\tilde b}$.  We  
introduce the momentum-space outer ellipsoid $\mathbb P$, which is the set of points $p$, such that
$bp_{L}^{2}+p_{\perp}^{2}<\Lambda^{2}$. We define a second, inner ellipsoid 
${\tilde {\mathbb P}}$ 
to
be the
set of points $p$, such that ${\tilde b}p_{L}^{2}+p_{\perp}^{2}<{\tilde \Lambda}^{2}$. Finally,
we define $\mathbb S$ to be the complement of ${\tilde {\mathbb P}}$ in $\mathbb P$. This is the shell or ``onion skin" between the two ellipsoidal surfaces:
 ${\mathbb S}={\mathbb P}-{\tilde{\mathbb P}}$. 

We split our fields into ``slow" and ``fast" pieces:
\beq
\psi(x)=\tilde{\psi}(x)+\varphi (x), \;
\bar{\psi}(x)= \tilde{\bar{\psi}}(x)+\bar{\varphi}(x),\;
A_{\mu}(x) =\tilde{A_{\mu}} (x)+a_{\mu} (x), \label{slow-fast}
\eeq
where the Fourier components of $\psi(x)$, ${\bar \psi}(x)$ and $A_{\mu}(x)$ vanish outside the
ellipsoid $\mathbb P$, the Fourier components of the slow fields  ${\tilde \psi}(x)$, ${\tilde{ \bar{ \psi}}}(x)$ and ${\tilde A}_{\mu}(x)$ vanish outside the
inner ellipsoid $\tilde {\mathbb P}$, and the Fourier components of the fast fields 
$\varphi(x)$, ${\bar \varphi}(x)$ and $a_{\mu}(x)$ vanish outside of the shell $\mathbb S$. Explicitly,
\beq
\tilde{\psi}(x)=\int_{\tilde{\mathbb{P}}}\frac{d^{4}p}{(2\pi)^{4}}\psi(p)e^{-ip\cdot x}, \,\,\varphi(x)
\!&\!=\!&\!\int_{\mathbb{S}}\frac{d^{4}p}{(2\pi)^{4}}\psi(p)e^{-ip\cdot x},\;
\tilde{\bar{\psi}}(x)=\int_{\tilde{\mathbb{P}}}\frac{d^{4}p}{(2\pi)^{4}}\bar{\psi}(p)e^{ip\cdot x},\,\,\bar{\varphi}(x)=\int_{\mathbb{S}}\frac{d^{4}p}{(2\pi)^{4}}\bar\psi(p)e^{ip\cdot x}, \nonumber \\
\tilde{A}_{\mu}(x)\!&\!=\!&\!
\int_{\tilde{\mathbb{P}}}\frac{d^{4}p}{(2\pi)^{4}}A_{\mu}(p)e^{-ip\cdot x},\,\,\,a_{\mu}(x)=\int_{\mathbb{S}}\frac{d^{4}p}{(2\pi)^{4}}A_{\mu}(p)e^{-ip\cdot x}.\nonumber
\eeq

The covariant derivative becomes 
\beq
D_\mu=\partial_\mu-i\tilde{A}_\mu-ia_\mu=\tilde{D}_\mu-ia_\mu, \nonumber       
\eeq
 where $\tilde{D}_\mu$ is the ``slow" covariant derivative. The field strength is 
 \beq
 F_{\mu\nu}=\tilde{F}_{\mu\nu}+[\tilde{D}_\mu,a_\nu]-[\tilde{D}_\nu, a_\mu]-i[a_\mu, a_\nu], \nonumber
 \eeq
where $\tilde{F}_{\mu\nu}$ is the slow field strength 
$\tilde{F}_{\mu\nu}=i[\tilde{D}_\mu,\tilde{D}_\nu]$. The Yang-Mills action is
\beq
S_G=\int d^4x\frac{1}{4g_0^2}\left(\tilde{F}_{\mu\nu}\tilde{F}^{\mu\nu}-4[\tilde{D}_\mu,\tilde{F}^{\mu\nu}]a_\nu+\left([\tilde{D}_\mu,a_\nu]-[\tilde{D}_\nu, a_\mu]\right)\left([\tilde{D}^\mu, a^\nu]-[\tilde{D}^\nu, a^\mu]\right)-2i\tilde{F}^{\mu\nu}[a_\mu,a_\nu]\right),\nonumber
\eeq
to quadratic order in $a_\mu$.

To do perturbation theory, we add a gauge-fixing term $\frac{1}{2g_0^2}\int d^4x {\rm Tr} [\tilde{D}_\mu, a_\mu]^2$ to the action. This reduces the gauge symmetry of the fast fields, and means that we must also introduce Faddeev-Popov ghost fields. The Yang-Mills action becomes
\beq
S_G=\frac{1}{4g_0^2}\int d^4x \tilde{F}_{\mu\nu}\tilde{F}^{\mu\nu}+\frac{1}{2g_0^2}\int d^4x\left([\tilde{D}_\mu,a_\nu][\tilde{D}^\mu, a^\nu]-2i\tilde{F}^{\mu\nu}[a_\mu, a_\nu]\right)\nonumber.
\eeq
The expansion of the quark-field action into slow and fast components is
\beq
S_{\rm Dirac}=\int d^4x\left(\tilde{\bar{\psi}}\,/\!\!\!\!\partial\,\tilde{\psi}+\bar{\varphi}\,/\!\!\!\!\partial\,\varphi+\bar{\varphi}\,/\!\!\!\!\tilde{A}\,\varphi+\tilde{\bar{\psi}}\,/\!\!\!\!a\,\varphi+\bar{\varphi}\,/\!\!\!\!a\,\tilde{\psi}\right).\nonumber
\eeq
The action is that of free slow and fast fields plus interaction terms
\beq
S=\tilde{S}+S_0+S_{I}+S_{II}+S_{1}+S_{2}+S_{3}+S_{\rm ghost} \nonumber,
\eeq
where
\beq
\tilde{S}&=&\frac{1}{4g_0^4}\int d^4x\tilde{F}_{\mu\nu}\tilde{F}^{\mu\nu}+\int d^4x \tilde{\bar{\psi}}\,/\!\!\!\!\partial\tilde{\psi},\nonumber\\
S_0&=&\frac{1}{2g_0^2}\int_{\mathbb{S}}\frac{d^4q}{(2\pi)^2}q^2a_\mu^b(-q)a^{\mu\, b}(q)-i\int_\mathbb{S}\frac{d^4q}{(2\pi)^4}\bar{\varphi}(-q)/\!\!\!q\varphi(q)\nonumber,\\
S_I&=&\frac{i}{2g_0^2}\int_\mathbb{S}\frac{d^4q}{(2\pi)^4}\int_{\tilde{\mathbb{P}}}\frac{d^4p}{(2\pi)^4}q^\mu f_{bcd}a_\nu^b(q)\tilde{A}_\mu^c(p)a^{\nu\,d}(-q-p)\nonumber\\
&+&\frac{1}{2g_0^2}\int_\mathbb{S}\frac{d^4q}{(2\pi)^4}\int_{\tilde{\mathbb{P}}}\frac{d^4p}{(2\pi)^4}\int_{\tilde{\mathbb{P}}}\frac{d^4l}{(2\pi)^4}f_{bcd}f_{bfg}a_\nu^d(q)\tilde{A}_\mu^c(p)\tilde{A}^{\mu\,f}(l)a^{\nu\,g}(-q-p-l),\nonumber\\
S_{II}&=&\frac{1}{2g_0^2}\int_\mathbb{S}\frac{d^4q}{(2\pi)^4}\int_{\tilde{\mathbb{P}}}\frac{d^4p}{(2\pi)^4}f_{bcd}a_\mu^b(q)\tilde{F}^{\mu\nu\,c}(p)a_\nu^d(-p-q),\nonumber\\
S_1&=&\int_\mathbb{S}\frac{d^4q}{(2\pi)^4}\int_{\tilde{\mathbb{P}}}\frac{d^4p}{(2\pi)^4} \bar{\varphi}(p)/\!\!\!\!A(q)\varphi(-q-p),\nonumber\\
S_2&=&\int_\mathbb{S}\frac{d^4q}{(2\pi)^4}\int_{\tilde{\mathbb{P}}}\frac{d^4p}{(2\pi)^4}\tilde{\bar{\psi}}(p)/\!\!\!\!\,a(q)\varphi(-q-p),
\nonumber\\
S_3&=&S_2^*=\int_\mathbb{S}\frac{d^4q}{(2\pi)^4}\int_{\tilde{\mathbb{P}}}\frac{d^4p}{(2\pi)^4}\bar{\varphi}(-q-p)\,/\!\!\!\!a(q)\varphi(p)\nonumber 
\eeq
and the ghost action, needed to have the correct measure on the fast field, is 
\beq
S_{\rm ghost}&=&\frac{i}{g_0^2}\int_\mathbb{S}\frac{d^4q}{(2\pi)^4}\int_{\tilde{\mathbb{P}}}\frac{d^4p}{(2\pi)^4}q^\mu f_{bcd}G^b(q)\tilde{A}_\mu^c(p)H^d(-q-p)\nonumber\\
&+&\frac{1}{2g_0^2}\int_\mathbb{S}\frac{d^4q}{(2\pi)^4}\int_{\tilde{\mathbb{P}}}\frac{d^4p}{(2\pi)^4}\int_{\tilde{\mathbb{P}}}\frac{d^4l}{(2\pi)^4}f_{bcd}f_{bfg}G^d(q)\tilde{A}_\mu^c(p)\tilde{A}_\mu^f(l)H^g(-q-p)\nonumber.
\eeq
The interaction is therefore $S_{\rm int}=S_{I}+S_{II}+S_{1}+S_{2}+S_{3}+S_{\rm ghost}$.

In Wilsonian renormalization, we start with the functional integral
\beq
Z=\int_{\tilde{\mathbb{P}}}\mathcal{D}\tilde{\psi}\mathcal{D}\tilde{\bar{\psi}}\mathcal{D}\tilde{A}\,e^{-\tilde{S}}\int_\mathbb{S}\mathcal{D}\varphi\mathcal{D}\bar{\varphi}\mathcal{D}a \,e^{-S_0-S_{\rm int}},\label{functionalintegral}
\eeq
and integrate out the fast fields $\varphi$, $\bar \varphi$ and $a$, to obtain an action $S'$, defined by
\beq
e^{-S'}=
e^{-\tilde{S}}\int_{\mathbb{S}}\mathcal{D}\varphi\mathcal{D}\bar{\varphi}\mathcal{D}a\,e^{-S_0-S_{\rm int}}.\label{effectiveAc}
\eeq
Then the functional integral in terms of slow degrees of freedom only is 
\beq
Z=\int_{\tilde{\mathbb{P}}}\mathcal{D}\tilde{\psi}\mathcal{D}\tilde{\bar{\psi}}\mathcal{D}\tilde{A}e^{-S'}.\nonumber
\eeq
The Green's functions of the slow fields $\psi$, $\bar \psi$ and $\tilde A$, with action $S^{\prime}$ are unchanged from the same Green's functions in the original theory.

The effective action is given explicitly by $S^{\prime}={\tilde S}-\ln\langle e^{-S_{\rm int}}\rangle$, where we define the brackets for any observable $Q$ by
\beq
\langle Q\rangle=\left(\int_\mathbb{S}\mathcal{D}\varphi\mathcal{D}\bar{\varphi}\mathcal{D}a\,e^{-S_0}\right)^{-1}\int_\mathbb{S}\mathcal{D}\varphi\mathcal{D}\bar{\varphi}\mathcal{D}a\,Q\,e^{-S_0}.\nonumber
\eeq
To find $S^{\prime}$, we use the connected-graph expansion:
\beq
\langle e^{-S_{\rm int}}\rangle=\exp\left[-\langle S_{\rm int}\rangle+\frac{1}{2}\left(\langle S_{\rm int}^2\rangle-\langle S_{\rm int}\rangle^2\right)-\frac{1}{3!}\left(\langle S_{\rm int}^3\rangle-3\langle S_{\rm int}^2\rangle\langle S_{\rm int}\rangle+\langle S_{\rm int}\rangle^3\right)+...\right].\nonumber
\eeq 
To evaluate $\langle S_{\rm int}\rangle,\,\langle S_{\rm int}^2\rangle,$ and $\langle S_{\rm int}^3\rangle$, we need the fast-field propagators
\beq
\langle a^b_\mu(q)a^c_\nu(p)\rangle=g_0^2\delta^{bc}\delta_{\mu\nu}\delta^4(q+p)q^{-2}(2\pi)^4,\,\,\,\,\,\langle\varphi(p)\bar{\varphi}(q)\rangle=\frac{i\,/\!\!\!\!q}{q^2}\delta^4(p+q)(2\pi)^4.\nonumber
\eeq

We first consider all the contributions quadratic in the slow gauge field, {\em i.e.} vacuum polarization. One contribution comes from the interactions $S_I$ and $S_{\rm ghost}$:
\beq
&\langle S_I\rangle-\frac{1}{2}\left(\langle S_I^2\rangle-\langle S_I\rangle^2\right)+\langle S_{\rm ghost}\rangle-\frac{1}{2}\left(\langle S_{\rm ghost}^2\rangle-\langle S_{\rm ghost}\rangle^2\right)\nonumber\\
&=\frac{C_G}{4}\int_{\tilde{\mathbb{P}}}\frac{d^4p}{(2\pi)^4}\tilde{A}_\mu^b(-p)\tilde{A}_\nu^b(p)P_{\mu\nu}(p),\nonumber
\eeq
where
\beq
P_{\mu\nu}(p)=\int_\mathbb{S}\frac{d^4p}{(2\pi)^4}\left[-\frac{q_\mu(p_\nu+2q_\nu)}{4q^2(q+p)^2}+\frac{\delta_{\mu\nu}}{4q^2}\right],\nonumber
\eeq
and $C_G$ is the quadratic Casimir operator in the adjoint representation, defined by 
$C_G \delta^{bh}=f^{bcd}f^{hcd}$. The tensor $P_{\mu\nu}(p)$ is not symmetric under exchange of indices. We define the integral $I_{\alpha}(p)$ by
\beq
I_{\alpha}(p)=\int_\mathbb{S}\frac{d^4q}{(2\pi)^4}\frac{p_\alpha+2q_\alpha}{q^2(q+p)^2}\nonumber
\eeq
and notice that $I_{\alpha}(p)+I_{\alpha}(-p)=0$ (we can see this by changing the sign of $q$ in the integrand). We can then use this to  replace the tensor $P_{\mu\nu}(p)$ by the manifestly symmetric tensor $\Pi_{\mu\nu}^1(p)$:
\beq
&\langle S_I\rangle-\frac{1}{2}\left(\langle S_I^2\rangle-\langle S_I\rangle^2\right)+\langle S_{\rm ghost}\rangle-\frac{1}{2}\left(\langle S_{\rm ghost}^2\rangle-\langle S_{\rm ghost}\rangle^2\right)\nonumber\\
&=\int_{\tilde{\mathbb{P}}}\frac{d^4p}{(2\pi)^4}\tilde{A}_\mu^b(-p)\tilde{A}_\nu^b(p)\Pi^1_{\mu\nu}(p),\label{poltensorone}
\eeq
where
\beq
\Pi_{\mu\nu}^1(p)=C_G\int_\mathbb{S}\frac{d^4q}{(2\pi)^4}\left[-\frac{\left(p_\mu+2q_\nu\right)\left(p_\nu+2q_\mu\right)}{8q^2(p+q)^2}+\frac{\delta_{\mu\nu}}{4q^2}\right].\nonumber
\eeq
A second contribution to vacuum polarization comes from 
\beq
&-\frac{1}{2}\left(\langle S_{II}^2\rangle-\langle S_{II}\rangle^2\right)=-\frac{C_G}{2}\int_{\tilde{\mathbb{P}}}\frac{d^4p}{(2\pi)^2}\tilde{F}_{\mu\nu}^b(-p)\tilde{F}_{\mu\nu}^b(p)\int_\mathbb{S}\frac{d^4q}{(2\pi)^2}\frac{1}{q^2(p+q)^2}.\label{spinorbit}
\eeq
The third and final contribution to vacuum polarization comes from integration over the fast quark field:
\beq
-\frac{1}{2}\left(\langle S_1^2\rangle-\langle S_1^2\rangle^2\right)=\int_{\tilde{\mathbb{P}}}\frac{d^4p}{(2\pi)^2}\tilde{A}_\mu^b(-p)\tilde{A}_\nu^b(p)\Pi_{\mu\nu}^3(p),\label{vacuumfermion}
\eeq
where
\beq
\Pi_{\mu\nu}^3(p)=\frac{N_{f}}{2}\int_\mathbb{S}\frac{d^4q}{(2\pi)^2}{\rm Tr}\left[\frac{/\!\!\!\!q}{q^2}\gamma^\mu\frac{/\!\!\!\!q+/\!\!\!\!p}{(q+p)^2}\gamma^\nu\right],\nonumber
\eeq
and $N_{f}$ is the number of flavors. Combining (\ref{poltensorone}), (\ref{spinorbit}) and (\ref{vacuumfermion}), we find the vacuum-polarizaton contribution:
\beq
\int_{\tilde{\mathbb{P}}}\frac{d^4p}{(2\pi)^2}\tilde{A}^b_\mu(-p)\tilde{A}^b_\nu(p)\Pi_{\mu\nu}(p),\label{polarization}
\eeq
where
\beq
\Pi_{\mu\nu}(p)=\Pi_{\mu\nu}^1(p)+\Pi_{\mu\nu}^2(p)+\Pi_{\mu\nu}^3(p),\nonumber
\eeq
and
\beq
\Pi_{\mu\nu}^2(p)=(p^2\delta_{\mu\nu}-p_\mu p_\nu)C_G\int_\mathbb{S}\frac{-1}{2q^2(p+q)^2}.\nonumber
\eeq

The quark self-energy contribution, which comes from the interactions $S_2$ and $S_3$, is
\beq
-\frac{1}{2}\left(\langle S_2S_3\rangle+\langle S_3S_2\rangle\right)=\int_{\tilde{\mathbb{P}}}\frac{d^4p}{(2\pi)^2}\Sigma(p)\tilde{\bar{\psi}}(p)\tilde{\psi(p)},\label{selfenergy}
\eeq
where
\beq
\Sigma(p)=2g_0^2\int_\mathbb{S}\frac{d^4q}{(2\pi)^4}\left[\frac{i(/\!\!\!\!p+/\!\!\!\!q)}{q^2(p+q)^2}\right].\nonumber
\eeq

The quark-gluon vertex receives a correction from
\beq
&\frac{1}{3!}\left(\langle S_1S_2S_3\rangle-\langle S_2S_3\rangle\langle S_1\rangle-\langle S_3S_2\rangle \langle S_1\rangle\right)+\frac{1}{3!}\left(\langle S_IS_2S_3\rangle-\langle S_2S_3\rangle\langle S_I\rangle-\langle S_3S_2\rangle \langle S_I\rangle\right)\nonumber\\
&=\int_{\tilde{\mathbb{P}}}\frac{d^4q}{(2\pi)^4}\int_{\tilde{\mathbb{P}}}\frac{d^4p}{(2\pi)^4}\tilde{\bar{\psi}}(p)\Gamma^{\mu\,a}(p,q)\tilde{A}_\mu^a(q)\tilde{\psi}(-q-p),\label{vertex}
\eeq
where
\beq
\Gamma^{\mu\,a}(p,q)=-2g_0^2 t^a\int_\mathbb{S}\frac{d^4k}{(2\pi)^4}\frac{/\!\!\!\!k \gamma^\mu(/\!\!\!\!k+/\!\!\!\!q)}{(k-p)^2(k+q)^2k^2}.\nonumber
\eeq

Finally, after $S^{\prime}$ is obtained, we rescale longitudinally, so that the ellipsoidal space $\tilde {\mathbb P}$ becomes a sphere. This will change $S^{\prime}$ into the effective action $S_{\rm eff}$ with a spherical cut-off.

\section{Renormalization with spherical cut-offs}

We can recover spherical cut-offs by setting $b=\tilde{b}=1$. This is the situation usually considered in treatments of Wilsonian renormalization, for $\tilde{\mathbb{P}}$ and $\mathbb{P}$ become spheres in momentum space. The set of momentum-space points $\mathbb{S}$ is the spherical 
shell, defined by $\mathbb{S}=\mathbb{P}-\tilde{\mathbb{P}}$.

We expand $\Pi_{\mu\nu}(p), \Sigma(p)$ and $\Gamma^{\mu\,a}(p,q)$ in powers of the slow momenta, treating momenta in $\tilde{\mathbb{P}}$ as much smaller than momenta in $\mathbb{S}$. This gives
\beq
&\Pi_{\mu\nu}(p)=\Pi^1_{\mu\nu}(p)+\Pi^2_{\mu\nu}(p)+\Pi^3_{\mu\nu}(p),\nonumber\\
&\Pi^1_{\mu\nu}(p)=C_G\left[\frac{\delta_{\mu\nu}}{4}E-\frac{1}{2}A_{\mu\nu}+\frac{p_\mu p_\alpha}{2}B_{\nu\alpha}+\frac{p_\nu p_\alpha}{2}B_{\nu\alpha}+\frac{p_\nu p_\alpha}{2}B_{\mu\alpha}-\frac{p_\mu p_\nu}{8}D\right.\nonumber\\
&\left.+\frac{p^2}{2}B_{\mu\nu}-2p_\alpha p_\beta C_{\alpha\beta\mu\nu}\right],\nonumber\\
&\Pi^2_{\mu\nu}(p)=-\frac{1}{2}(p^2g_{\mu\nu}-p_\mu p_\nu) C_G D,\nonumber\\
&\Pi^3_{\mu\nu}(p)=\frac{N_{f}}{2}{\rm Tr}\gamma^\mu\gamma^\alpha\gamma^\nu\gamma^\beta \left[A_{\alpha\beta}+4C_{\alpha\beta\gamma\delta}p^\gamma p^\delta-p^2B_{\alpha\beta}-2B_{\alpha\gamma}p_\beta p^\gamma\right],\nonumber\\
&\Sigma(p)=2g_0^2N_{f}\gamma^\alpha\left[-2B_{\alpha\beta}p^\beta+p_\alpha D\right],\nonumber\\
&\Gamma^{\mu\,a}=-2g_0^2N_{f}t^a\gamma^\alpha\gamma^\mu\gamma^\beta B_{\alpha\beta},\label{spherical}
\eeq
where
\beq
&A_{\alpha\beta}=\int_\mathbb{S}\frac{d^4q}{(2\pi)^4}\frac{q_\alpha q_\beta}{q^4},\,\,\,\,\,B_{\alpha\beta}=\int_\mathbb{S}\frac{d^4q}{(2\pi)^4}\frac{q_\alpha q_\beta}{q^6},\nonumber\\
&C_{\alpha\beta\gamma\delta}=\int_\mathbb{S}\frac{d^4q}{(2\pi)^4}\frac{q_\alpha q_\beta q_\gamma q_\delta}{q^8},\,\,\,\,\,D=\int_\mathbb{S}\frac{d^4q}{(2\pi)^4}\frac{1}{q^4},\,\,\,\,\,E=\int_\mathbb{S}\frac{d^4q}{(2\pi)^4}\frac{1}{q^2}.\label{defintegrals}
\eeq
The integrals (\ref{defintegrals}) are invariant under $\mathcal{O}(4)$ rotation symmetry. This allows us to write
\beq
\int_\mathbb{S}d^4q\,q_\alpha q_\beta=\frac{\pi^2}{2}\int_{\tilde{\Lambda}}^\Lambda dq\,
\delta_{\alpha\beta}\,q^{2}\;, \label{ofoursymmetryone}
\eeq
and
\beq
\int_{\mathbb{S}}d^4q\,q_\alpha q_\beta q_\gamma q_\delta=\frac{1}{24}\int_\mathbb{S} d^4q\, q^4 (\delta_{\alpha\beta} \delta_{\gamma\delta}+\gamma_{\alpha\delta}\delta_{\gamma\beta}+\delta_{\alpha\gamma}\delta_{\beta\delta}).\label{ofoursymmetrytwo}
\eeq

Using (\ref{ofoursymmetryone}) and (\ref{ofoursymmetrytwo}) we solve (\ref{spherical}):
\beq
&\Pi_{\mu\nu}(p)=-\frac{11C_G}{192\pi^2}(p^2\delta_{\mu\nu}-p_\mu p_\nu)\ln\frac{\Lambda}{\tilde{\Lambda}} +\frac{N_{f}}{12\pi^2}(p^2\delta_{\mu\nu}-p_\mu p_\nu)\ln\frac{\Lambda}{\tilde{\Lambda}}\nonumber\\
&+\frac{C_G}{128\pi^2}(\Lambda^2-\tilde{\Lambda}^2)\delta_{\mu\nu}-\frac{N_{f}}{16\pi^2}(\Lambda^2-\tilde{\Lambda})\delta_{\mu\nu},\nonumber\\
&\Sigma(p)=g_0^2 N_{f} \frac{\gamma^\mu p_\mu}{8\pi^2}\ln\frac{\Lambda}{\tilde{\Lambda}},\nonumber\\
&\Gamma^{a\mu}=g_0^2 N_{f} t^a\frac{\gamma^\mu}{8\pi^2}\ln\frac{\Lambda}{\tilde{\Lambda}}.\label{resultsspherical}
\eeq

The terms in the polarization tensor that are quadratic in the cut-offs produce corrections to the action that break gauge invariance. We can fix this problem by introducing mass counterterms in the action at each scale to cancel these. We keep the gauge invariant part of the polarization tensor, which we call $\hat{\Pi}_{\mu\nu}(p)$, and is defined by
\beq
\hat{\Pi}_{\mu\nu}(p)=\Pi_{\mu\nu}(p)-\Pi_{\mu\nu}(0).\nonumber
\eeq

The resulting action for the slow fields has the coupling $\tilde{g}$, given by
\beq
\frac{1}{4\tilde{g}^2}=\frac{1}{4g_0^2}-\frac{11C_G}{192\pi^2}\ln\frac{\Lambda}{\tilde{\Lambda}}+\frac{N_{f}}{12\pi^2}\ln\frac{\Lambda}{\tilde{\Lambda}}, \label{effectivecoupling}
\eeq
which is the standard result.

\section{Renormalization with ellipsoidal cut-offs}

In this section, we generalize to the case where the region 
$\mathbb{S}$ is an ellipsoidal shell. We first change from the four variables $q_\mu \in \mathbb{S}$, to two radial variables (with units of momentum squared) $u,$ and $w$, and two angles, $\theta$ and $\phi$. We write the old variables in terms of these new variables by
\beq
q_1=\sqrt{u}\cos \theta,\,\,q_2=\sqrt{u}\sin\theta,\,\,q_3=\sqrt{w-u}\cos\phi,\,\,q_0=\sqrt{w-u}\sin\phi.\nonumber
\eeq
In terms of the new variables, the integration over $\mathbb{S}$ is
\beq
\int_{\mathbb{S}} d^4q=\frac{1}{4}\int_0^{2\pi}d\theta\int_0^{2\pi}d\phi\left[\int_0^{\tilde{\Lambda}^2}du\int_{\tilde{b}^{-1}\tilde{\Lambda}^2+(1-\tilde{b}^{-1})u}^{b^{-1}\Lambda^2+(1-b^{-1})u}dw+\int_{\tilde{\Lambda}^2}^{\Lambda^2}du\int_u^{b^{-1}\Lambda^2+(1-b^{-1})u}dw\right].\label{newvariables}
\eeq
There is no longer an $\mathcal{O}(4)$ symmetry of the integration measure. There is now only the symmetry $\mathcal{O}(2)\times\mathcal{O}(2)$, generated by the rotations $\theta\to\theta+d\theta$ and $\phi\to\phi+d\phi$. We also introduce some new notation to distinguish transverse from longitudinal Lorentz indices. The letters $C$ and $D$ denote Lorentz indices with values 1 and 2. The letters $\Omega$ and $\Xi$ denote Lorentz indices with values 0 and 3. With this 
preparation, we are ready to carry out the necessary integration. 

The integrals (\ref{defintegrals}) written in terms of the new variables (\ref{newvariables}), reduce to
\beq
A_{CD}&=&\frac{\delta_{CD}}{32\pi^2}\Lambda^2\left[1+\frac{b}{(b-1)^2}(1-b+\ln b)\right]-\frac{\delta_{CD}}{32\pi^2}\tilde{\Lambda}^2\left[1+\frac{\tilde{b}}{(\tilde{b}-1)^2}(1-\tilde{b}+\ln\tilde{b})\right],\nonumber\\
A_{\Omega\Xi}&=&\frac{\delta_{\Omega\Xi}}{32\pi^2}\left\{\Lambda^2\left[\frac{1}{b-1}-\frac{\ln b}{(b-1)^2}\right]-\tilde{\Lambda}^2\left[\frac{1}{\tilde{b}-1}-\frac{\ln\tilde{b}}{(\tilde{b}-1)^2}\right]\right\},\nonumber\\
A_{C\Omega}&=&0,\nonumber\\
B_{CD}&=&\frac{\delta_{CD}}{32\pi^2}\ln\frac{\Lambda}{\tilde{\Lambda}}-\frac{\delta_{CD}}{64\pi^2}\left[\frac{b^2\ln b}{(b-1)^2}-\frac{b}{b-1}\right]+\frac{\delta_{CD}}{64\pi^2}\left[\frac{\tilde{b}^2\ln\tilde{b}}{(\tilde{b}-1)^2}-\frac{\tilde{b}}{\tilde{b}-1}\right],\nonumber\\
B_{\Omega\Xi}&=&\frac{\delta_{\Omega\Xi}}{32\pi^2}\ln\frac{\Lambda}{\tilde{\Lambda}}-\frac{\delta_{\Omega\Xi}}{64\pi^2}\left[\frac{b(b-2)\ln b}{(b-1)^2}+\frac{b}{b-1}\right]+\frac{\delta_{\Omega\Xi}}{64\pi^2}\left[\frac{\tilde{b}(\tilde{b}-2)\ln \tilde{b}}{(\tilde{b}-1)^2}+\frac{\tilde{b}}{\tilde{b}-1}\right],\nonumber\\
B_{C\Omega}&=&0,\nonumber\\
C_{CCCC}&=&\frac{1}{64\pi^2}\ln\frac{\Lambda}{\tilde{\Lambda}}-\frac{1}{128\pi^2}\frac{b^3}{(b-1)^3}\left[\ln b-\frac{2(b-1)}{b}+\frac{(b-1)(b+1)}{2b^2}\right]\nonumber\\
&+&\frac{1}{128\pi^2}\frac{\tilde{b}^3}{(\tilde{b}-1)^3}\left[\ln\tilde{b}-\frac{2(\tilde{b}-1)}{\tilde{b}}
+\frac{(\tilde{b}-1)(\tilde{b}+1)}{2\tilde{b}^2}\right],\nonumber\\
C_{1122}&=&\frac{1}{3}C_{CCCC}\nonumber\\
C_{\Omega\Omega\Omega\Omega}&=&\frac{1}{64\pi^2}\ln\frac{\Lambda}{\tilde{\Lambda}}
-\frac{1}{128\pi^2}\left\{\frac{b^3}{(b-1)^3}\left[\ln b-\frac{2(b-1)}{b}+\frac{(b-1)(b+1)}{2b^2}\right]+
\frac{3b\ln b}{b-1}-\frac{3b^2\ln b}{(b-1)^2}
+\frac{3b}{b-1}\right\}\nonumber\\
&+&\frac{1}{128\pi^2}\left\{\frac{\tilde{b}^3}{(\tilde{b}-1)^3}\left[\ln\tilde{b}-\frac{2(\tilde{b}-1)}{\tilde{b}}+\frac{(\tilde{b}-1)(\tilde{b}+1)}{2\tilde{b}^2}\right]+\frac{3\tilde{b}\ln\tilde{b}}{\tilde{b}-1}-\frac{3\tilde{b}\ln\tilde{b}}{(\tilde{b}-1)^2}+\frac{3\tilde{b}}{\tilde{b}-1}\right\},\nonumber\\
C_{0033}&=&\frac{C_{\Omega\Omega\Omega\Omega}}{3},\nonumber\\
C_{CC\Omega\Omega}&=&\frac{1}{192\pi^2}\ln\frac{\Lambda}{\tilde{\Lambda}}-\frac{1}{384\pi^2}\left[\frac{-2b\ln b}{(b-2)^3}+\frac{b^2\ln b+2b}{(b-1)^2}\right]+\frac{1}{384\pi^2}\left[\frac{-2\tilde{b}\ln\tilde{b}}{(\tilde{b}-1)^3}+\frac{\tilde{b}\ln \tilde{b}+2\tilde{b}}{(\tilde{b}-1)^2}\right],\nonumber\\
D&=&\frac{1}{8\pi^2}\ln\frac{\Lambda}{\tilde{\Lambda}}-\frac{1}{16\pi^2}\left[\frac{b\ln b}{b-1}-\frac{\tilde{b}\ln\tilde{b}}{\tilde{b}-1}\right],\nonumber\\
E&=&\frac{1}{16\pi^2}\left(\frac{\Lambda^2\ln b}{b-1}-\frac{\tilde{\Lambda}^2\ln\tilde{b}}{\tilde{b}-1}\right).\label{resultsellipsoid}
\eeq
If we substitute $b=\tilde{b}=1$ into (\ref{resultsellipsoid}), we recover (\ref{resultsspherical}) in the last section.

We take $b=1$ and $\tilde{b}\approx 1$. We expand $\tilde{b}=1+\ln\tilde{b}+\frac{\ln^2\tilde{b}}{2!}+\cdots$ and $\ln\tilde{b}=\ln\tilde{b}-\frac{\ln^2\tilde{b}}{2}+\cdots$, keeping only the first-order terms in $\ln\tilde{b}$.

The self-energy correction is
\beq
\Sigma(p)=2g_0^2\left[\frac{\gamma^\mu p_\mu}{16\pi^2}\ln\frac{\Lambda}{\tilde{\Lambda}}+\frac{1}{32\pi^2}\ln\tilde{b}\frac{\gamma^Cp_C}{6}-\frac{1}{32\pi^2}\ln\tilde{b}\frac{\gamma^\Omega p_\Omega}{6}\right].\label{selfenergyellipsoid}
\eeq
The vertex correction is
\beq
\Gamma^{\mu\,a}=-2g_0^2 t^a\left[\frac{-\gamma^\mu}{16\pi^2}\ln\frac{\Lambda}{\tilde{\Lambda}}-
\frac{\gamma^\mu}{16\pi^2}\ln\tilde{b}+\frac{g^{C\mu}\gamma_C}{32\pi^2}\frac{5}{6}\ln\tilde{b}+\frac{g^{\Omega\mu}\gamma_\Omega}{32\pi^2}\frac{7}{6}\ln\tilde{b}\right].\label{vertexellipsoid}
\eeq

The general form of the quadratic part of the renormalized gauge field action, which is invariant under $\mathcal{O}(2)\times\mathcal{O}(2)$ and gauge symmetry is
\beq
S_{\rm quadratic}=\int_{\tilde{\mathbb{P}}}\frac{d^4p}{(2\pi)^4}A(-p)^T[a_1M_1(p)+a_2M_2(p)+a_3M_3(p)]A(p),\nonumber
\eeq
where
\beq
&M_1(p)=\left(\begin{array}{cccc}p_2^2&-p_1p_2&0&0\\-p_1p_2&p_1^2&0&0\\0&0&0&0\\0&0&0&0\end{array}\right),\,\,\,\, M_2(p)=\left(\begin{array}{cccc}0&0&0&0\\0&0&0&0\\0&0&p_0^2&-p_3p_0\\0&0&-p_3p_0&p_3^2\end{array}\right),\nonumber\\
&M_3(p)=\left(\begin{array}{cccc}p_L^2&0&-p_1p_3&-p_1p_0\\0&p_L^2&-p_2p_3&-p_2p_0\\-p_1p_3&-p_2p_3&p_\perp^2&0\\-p_1p_0&-p_2p_0&0&p_\perp^2\end{array}\right),\label{mmatrices}
\eeq
and the coefficients $a_1,\,a_2$ and $a_3$ are real numbers. We extract these coefficients from the polarization tensor. Any part that cannot be expressed in terms of (\ref{mmatrices}) (i.e.  $S_{\rm diff}=\int_{\tilde{\mathbb{P}}}\frac{d^4p}{(2\pi)^4}A_\mu(-p)\Pi_{\mu\nu}(p)A_\nu(p)-S_{\rm quadratic})$ must be removed with counterterms in the action. The coefficients $a_i$ are selected such that $S_{\rm diff}$ is maximally non-gauge invariant. We insert (\ref{resultsellipsoid}) into (\ref{spherical}) and after some work we find
\beq
a_1&=&-\frac{11C_G}{192\pi^2}\ln\frac{\Lambda}{\tilde{\Lambda}}+\frac{N_{f}}{12\pi^2}\ln\frac{\Lambda}{\tilde{\Lambda}}-\frac{1}{64\pi^2}\frac{31}{9}C_G\ln\tilde{b}+\left(\frac{5}{48\pi^2}-\frac{1}{128\pi^2}\frac{104}{9}\right)N_{f}\ln\tilde{b},\nonumber\\
a_2&=&-\frac{11C_G}{192\pi^2}\ln\frac{\Lambda}{\tilde{\Lambda}}+\frac{N_{f}}{12\pi^2}\ln\frac{\Lambda}{\tilde{\Lambda}}-\frac{1}{64\pi^2}\frac{67}{9}C_G\ln\tilde{b}+\left(\frac{5}{48\pi^2}+\frac{1}{128\pi^2}\frac{40}{9}\right)N_{f}\ln\tilde{b},\nonumber\\
a_3&=&-\frac{11C_G}{192\pi^2}\ln\frac{\Lambda}{\tilde{\Lambda}}+\frac{N_{f}}{12\pi^2}\ln\frac{\Lambda}{\tilde{\Lambda}}-\frac{1}{64\pi^2}\frac{59}{9}C_G\ln\tilde{b}+\left(\frac{5}{48\pi^2}+\frac{1}{128\pi^2}\frac{8}{9}\right)\ln\tilde{b},\label{anumbers}
\eeq
and
\beq
M_{\rm diff}\!\!\!&\!\!\!=& \nonumber \\
&\frac{C_G\ln\tilde{b}}{64\pi^2}&\left(\begin{array}{cc}-\frac{1}{12}p_1^2-\frac{1}{2}p_2^2+\frac{7}{12}p+L^2&0\\0&-\frac{1}{2}p_1^2-\frac{1}{12}p_2^2+\frac{7}{12}p_L^2\\0&0\\0&0\end{array}
\begin{array}{cc}0&0\\0&0\\ \frac{7}{12}p_\perp^2+\frac{17}{12}p_3^2+\frac{5}{6}p_0^2&0\\0& \frac{7}{12}p_\perp^2+\frac{5}{6}p_3^2+\frac{17}{12}p_0^2\end{array}\right)\nonumber\\
& &\nonumber \\
& &\nonumber \\
&+&\frac{N_{f}\ln\tilde{b}}{128\pi^2}\frac{8}{3}\left(\begin{array}{cccc}\frac{17}{6}p_\perp^2+\frac{4}{3}p_L^2&0&0&0\\0&\frac{17}{6}p_\perp^2+\frac{4}{3}p_L^2&0&0\\0&0&-\frac{7}{6}p_L^2-\frac{14}{3}p_\perp^2&0\\0&0&0&-\frac{7}{6}p_L^2-\frac{14}{3}p_\perp^2\end{array}\right),\nonumber
\eeq
where $M_{\rm diff}$ is defined as
\beq
S_{\rm diff}=\int_{\tilde{\mathbb{P}}}\frac{d^4p}{(2\pi)^4}A(-p)^TM_{\rm diff}A(p).\nonumber
\eeq

\section{The renormalized action}

We next put together the results of the previous section to obtain the action 
$S'$, defined in (\ref{effectiveAc}). 
This action is 
\beq
S'=\int d^4x(\mathcal{L}_{\rm quarks}+\mathcal{L}_{\rm vertex}+\mathcal{L}_{\rm gauge}]=\int d^4x[\mathcal{L}_{\rm Dirac}+\mathcal{L}_{\rm gauge}),\nonumber
\eeq
where to one loop,
\beq
&\mathcal{L}_{\rm quarks}=\tilde{\bar{\psi}}i(/\!\!\!\partial+\Sigma(\partial))\tilde{\psi},\nonumber\\
&\mathcal{L}_{\rm vertex}=\tilde{\bar{\psi}}(\gamma^\mu t^a+\Gamma^{\mu\,a})\tilde{A}_\mu^a\tilde{\psi},\nonumber
\eeq
and
\beq
\mathcal{L}_{\rm gauge}=\frac{1}{4g_0^2}\tilde{F}_{\mu\nu}\tilde{F}^{\mu\nu}+\tilde{A}_\mu\left[\sum_{i=1}^3 a_iM_i^{\mu\nu}(\partial)\right]A_\nu.\label{lagrangians}
\eeq
Substituting (\ref{vertexellipsoid}) into (\ref{lagrangians}) yields
\beq
&\mathcal{L}_{\rm vertex}=R\tilde{\bar{\psi}}\left[\gamma^C\left(1+\frac{N_{f}g_0^2}{8\pi^2}\ln \frac{\Lambda}{\tilde{\Lambda}}+\frac{N_{f}g_0^2}{8\pi^2}\ln\tilde{b}-\frac{N_{f}5g_0^2}{96\pi^2}
\ln\tilde{b}\right)\right.A_C\nonumber\\
&\left.+\gamma^\Omega\left(1+\frac{N_{f}g_0^2}{8\pi^2}\ln\frac{\Lambda}{\tilde{\Lambda}}+\frac{N_{f}g_0^2}{8\pi^2}\ln\tilde{b}-\frac{N_{f}7g_0^2}{96\pi^2}\ln\tilde{b}\right)A_\Omega\right]\tilde{\psi}\nonumber
=R\tilde{\bar{\psi}}\left(\gamma^CA_C+\lambda^{\frac{N_{f}g_0^2}{24\pi^2\tilde{R}}}\gamma^\Omega A_\Omega\right)\tilde{\psi},\nonumber
\eeq
where
\beq
&R=\tilde{R}+N_{f}\left(\frac{g_0^2}{8\pi^2}-\frac{5g_0^2}{96\pi^2}\right)\ln\tilde{b}\approx\tilde{R}\tilde{b}^{\frac{7N_{f}g_0^2}{96\pi^2\tilde{R}}}=\tilde{R}\lambda^{-\frac{7N_{f}g_0^2}{48\pi^2\tilde{R}}},\nonumber\\
&\tilde{R}=1+\frac{N_{f}}{8\pi^2}g_0^2\ln\frac{\Lambda}{\tilde{\Lambda}},\label{defr}
\eeq
and where we have identified $\tilde{b}=\lambda^{-2}$ and dropped terms of order 
$(\ln\tilde{b})^{2}$.

We substitute (\ref{selfenergyellipsoid}) into (\ref{lagrangians}) to find
\beq
&\mathcal{L}_{\rm quarks}=\tilde{\bar{\psi}}\left[\gamma^C\partial_C\left(1+\frac{N_{f}g_0^2}{8\pi^2}\ln\frac{\Lambda}{\tilde{\Lambda}}+\frac{N_{f}g_0^2}{8\pi^2}\ln\tilde{b}-\frac{5N_{f}g_0^2}{96\pi^2}\ln\tilde{b}-\frac{N_{f}g_0^2}{16\pi^2}\ln\tilde{b}\right)\right.\nonumber\\
&\left.+\gamma^\Omega\partial_\Omega\left(1+\frac{N_{f}g_0^2}{8\pi^2}\ln\frac{\Lambda}{\tilde{\Lambda}}+\frac{N_{f}g_0^3}{8\pi^2}\ln\tilde{b}-\frac{5N_{f}g_0^2}{96\pi^2}\ln\tilde{b}-\frac{N_{f}g_0^2}{12\pi^2}\ln\tilde{b}\right)\right]\tilde{\psi}\nonumber\\
&\approx R'\tilde{\bar{\psi}}\,i\left[\gamma^C\partial_C+\lambda^{\frac{N_{f}g_0^2}{24\pi^2\tilde{R}}}\gamma^\Omega\partial_\Omega\right]\tilde{\psi},\nonumber
\eeq
where
\beq
R'=R\lambda^{\frac{N_{f}g_0^2}{8\pi^2\tilde{R}}}.\label{defrprime}
\eeq

To make the effective action manifestly gauge invariant, we need to $\mathcal{L}_{\rm Dirac}=\mathcal{L}_{\rm quarks}+\mathcal{L}_{\rm vertex}$ in terms of covariant derivatives. We accomplish this by redefining
\beq
\lambda^{-\frac{N_{f}g_0^2}{8\pi^2\tilde{R}}}\tilde{A}_\mu\to\tilde{A}_\mu,\,\,\,\,R'\lambda^{-1+\frac{N_{f}g_0^2}{24\pi^2\tilde{R}}}\tilde{\bar{\psi}}\tilde{\psi}\to\bar{\psi}\psi,\label{rescaleda}
\eeq
so that
\beq
\mathcal{L}_{\rm Dirac}=\bar{\psi}\,i\left(\lambda^{1-\frac{N_{f}g_0^2}{24\pi^2\tilde{R}}}\gamma^CD_C+\gamma^\Omega D_\Omega\right)\psi.\nonumber
\eeq
The factor absorbed by the gauge field in the rescaling (\ref{rescaleda}) modifies the pure gauge action. We notice that this factor depends on the effective coupling $\frac{g_0}{\tilde{R}}$ instead of the coupling $\tilde{g}$ from (\ref{effectivecoupling}). This is because the factor from (\ref{rescaleda}) arises from the quark self energy and the vertex corrections, instead of the vacuum polariztion.

Substituting (\ref{anumbers}) into (\ref{lagrangians}) gives us
\beq
&\mathcal{L}_{\rm gauge}=\frac{1}{4}\left(\frac{1}{g_0^2}-\frac{11}{48\pi^2}C_G\ln\frac{\Lambda}{\tilde{\Lambda}}+\frac{1}{12\pi^2}N_{f}\ln\frac{\Lambda}{\tilde{\Lambda}}-\frac{1}{64\pi^2}\frac{59}{9}C_G\ln\tilde{b}+\frac{1}{9\pi^2}N_{f}\ln\tilde{b}\right)(\tilde{F}_{01}^2+\tilde{F}_{02}^2+\tilde{F}_{13}^2+\tilde{F}_{23}^2)\nonumber\\
&+\frac{1}{4}\left(\frac{1}{g_0^2}-\frac{11}{48\pi^2}C_G\ln\frac{\Lambda}{\tilde{\Lambda}}+\frac{1}{12\pi^2}N_{f}\ln\frac{\Lambda}{\tilde{\Lambda}}-\frac{1}{64\pi^2}\frac{31}{9}C_G\ln\tilde{b}+\frac{1}{9\pi^2}N_{f}\ln\tilde{b}-\frac{7}{72\pi}N_{f}\ln\tilde{b}\right)\tilde{F}_{12}^2\nonumber\\
&\frac{1}{4}\left(\frac{1}{g_0^2}-\frac{11}{48\pi^2}C_G\ln\frac{\Lambda}{\tilde{\Lambda}}+\frac{1}{12\pi^2}N_{f}\ln\frac{\Lambda}{\tilde{\Lambda}}-\frac{1}{64\pi^2}\frac{67}{9}C_G\ln\tilde{b}+\frac{1}{9\pi^2}N_{f}\ln\tilde{b}+\frac{1}{36\pi^2}N_{f}\ln\tilde{b}\right)\tilde{F}_{03}^2.\nonumber
\eeq

The pure gauge Lagrangian is then
\beq
&\mathcal{L}_{\rm gauge}=\frac{1}{4g_{\rm eff}^2}\left(\tilde{F}_{01}^2+\tilde{F}_{02}^2+\tilde{F}_{13}^2+\tilde{F}_{23}^2+\tilde{F}_{03}^2\lambda^{\frac{C_G}{32\pi^2}\frac{8}{9}\tilde{g}^2-\frac{2N_{f}}{9\pi^2}\tilde{g}}+\tilde{F}_{12}^2\lambda^{-\frac{C_G}{32\pi^2}\frac{28}{9}\tilde{g}^2+\frac{7N_{f}}{9\pi^2}\tilde{g}^2}\right),\nonumber
\eeq
where
\beq
g_{\rm eff}^2=\tilde{g}^2\lambda^{-\frac{C_G}{32\pi^2}\frac{59}{9}\tilde{g}^2+\frac{8N_{f}}{9\pi^2}\tilde{g}^2+\frac{N_{f}g_0^2}{4\pi^2\tilde{R}}},\label{geff}
\eeq
and
\beq
\frac{1}{\tilde{g}^2}=\frac{1}{g_0^2}-\frac{11}{48\pi^2}C_G\ln\frac{\Lambda}{\tilde{\Lambda}}+\frac{N_{f}}{3\pi^2}\ln\frac{\Lambda}{\tilde{\Lambda}}.\nonumber
\eeq
The last term in the powers of $\lambda$ in (\ref{geff}) comes from the  redefinitions (\ref{rescaleda}).

\section{The longitudinally rescaled effective action}

Finally, we longitudinally rescale the slow fields, after removing the tildes, and Wick rotating back to real space. The effective action is given by $S_{\rm eff}=\int d^4x \,\mathcal{L}_{\rm eff}$, where
\beq
\mathcal{L}_{\rm eff}&=&\frac{1}{4g_{\rm eff}^{2}}\left(F_{01}^2+F_{02}^2-F_{13}^2-F_{23}^2+
\lambda^{-2+\frac{C_G}{36\pi^2}\tilde{g}^2-\frac{2N_{f}}{9\pi^2}\tilde{g}^2}F_{03}^{2}
-\lambda^{2-
\frac{7C_G}{72\pi^2}\tilde{g}^2+\frac{7N_{f}}{9\pi^2}\tilde{g}^2}F_{12}^{2}\right) \nonumber \\
&+& {\bar \psi}_{\alpha}\,i
\left(
\lambda^{1-\frac{N_{f}g_0^2}{12\pi^2\tilde{R}}}\gamma^CD_C+\gamma^\Omega D_\Omega\right)\psi_\alpha , \label{finalaction}
\eeq
where the label $\alpha=1,2,...,N_f$ denotes the flavors of quarks.

The action (\ref{finalaction}) has the same form of the classically rescaled action (\ref{classicalrescaling}) and 
(\ref{classicalrescalingquark}), but has anomalous powers of $\lambda$, as well as an effective coupling constant 
(\ref{geff}). For a small number of flavors of quarks, the effective coupling becomes strong in the 
high-energy limit.

\section{Conclusion}

We have determined how the QCD action changes under a small longitudinal rescaling, $\lambda\lesssim 1$.
The effective action (\ref{finalaction}) is valid only for small rescalings. This action gives an indication as to how the high-energy theory should look, though it is dangerous to take it too seriously. This is because for $\lambda\ll1$ quantum fluctuations 
become large, invalidating perturbation theory. This 
difficulty is similar to that of finding long-distance effects in Yang-Mills theory. Though it may be a good assumption that the high-energy
limit is described by a simple cut-off action with a large coupling, this assumption cannot be proved using only perturbation theory.  In particular, we do not know the role of 
non-renormalizable terms far from the ultraviolet-free fixed point, arising from the integration over high-energy components of fields. With these caveats, our result does seem to support
the use of longitudinally-rescaled QCD in high-energy scattering \cite{Verlinde-squared},
\cite{McLerranVenugopalan}, \cite{nearintegrability}.

Effective high-energy actions 
need to be explored further. In Reference
\cite{nearintegrability} shown that the longitudinally-rescaled Yang-Mills action is equivalent to a set of coupled $(1+1)$-dimensional principal chiral nonlinear sigma models. The particles of the sigma model are 
soliton-like excitations whose scattering matrix is exactly known. These excitations have the colors of a fundamental-antifundamental pair.
The theory has a mass gap and confines quarks in the longitudinally rescaled limit.  This model is
probably too naive however. As perturbation theory cannot tell the details of the high-energy action,
a more reasonable assumption is to replace the sigma model by a more general 
$(1+1)$-dimensional massive field theory, with some Lorentz-invariant current form factors. The mass gap and the form factors would be parameters in the effective theory.

We hope to do a gauge-invariant method to find the longitudinally-rescaled action. This may be possible with some version of dimensional regularization, where the number of longitudinal and transverse dimensions can vary independently.  The background-field method could be used  instead of Wilsonian renormalization. This would eliminate the need for counterterms to maintain gauge invariance.

\begin{acknowledgements}

We both thank the Niels Bohr International Academy and P.O.  thanks the Kavli Institute for Theoretical Physics, for their hospitality while we worked on this project. We also thank Jamal Jalilian-Marian for discussions. P.O.'s research was supported in part by the National Science Foundation, under Grant No. PHY0855387 and by a grant from the PSC-CUNY.

\end{acknowledgements}

\end{document}